\documentclass[12pt,preprint]{aastex}
\usepackage{graphicx}
\begin{document}
\title{Science-Operational Metrics and Issues\break for the ``Are We Alone?'' Movement}

\author{Robert A.\ Brown}
\affil{Space Telescope Science Institute}
\email{rbrown@stsci.edu
\vspace{12pt}}

A movement is underway to test the uniqueness of Earth. Sponsored primarily by NASA, it is enlisting talented researchers from many disciplines. It is conceiving new telescopes to discover and characterize other worlds like Earth around nearby stars and to obtain their spectra. The goal is to search for signs of biological activity and perhaps find other cradles of life. 

Most effort thus far has focused on the optics to make such observations feasible. Relatively little attention has been paid to science operations---the link between instrument and science. Because of the special challenges presented by extrasolar planets, science-operational issues may be limiting factors for the ``Are We Alone?''~(AWA) movement. Science-operational metrics can help compare the merits of direct and astrometric planet searches, and estimate the concatenated completeness of searching followed by spectroscopy. This completeness is the prime science metric of the AWA program. Therefore, the goals of this white paper are to present representative calculations involving science-operational metrics, and to promote a science-operational perspective. We urge the Survey Committee to allow this perspective and such metrics to inform its plan for the future of AWA.

\subsubsection*{Extreme targets}

Because of three features, an Earth twin (ET)---the AWA benchmark adopted for the calculations presented here---is the most extreme target in astronomy: low information rate, high variability in position and brightness, and the huge, proximate, unwanted signal of the star. To illustrate, consider a space telescope with aperture $D=16$~m, the largest currently under serious discussion. With efficiency $\varepsilon = 0.2$, and resolving power $R=10$ in the $I$~band, this telescope would detect 0.1 photons per second from an ET at quadrature at 20-pc distance. The apparent separation of the stellar and planetary images would be 0.05~arcsec or $5\, \lambda/D$. Assuming an inner working angle $\mathit{IWA}=3\,\lambda/D$---meaning the radius of the photometrically inaccessible region immediately around the star---and assuming a maximum feasible difference in magnitudes between the planet and star $\Delta mag_{0}=26$, set by wavefront instability for an internal coronagraph, or by the starshade accuracy and alignment in the case of an external occultor, such an ET on a random orbit would be detectable 66\% of random search epochs.\footnote{See calculations of completeness and discussion of $\Delta mag_{0}$ in Brown, R.~A. 2005, ``Single-Visit Photometric and Obscurational Completeness,'' \textit{ApJ}, 624, 1010--1024.} The median time for a detected ET on a partially obscured orbit to become undetectable again---too faint or too close to the star---is two months after detection. Assuming the optics suppress the diffraction wings by factor 10$^{-10}$ from the central intensity of the stellar image, and assuming sharpness $\Psi=0.02$, some 10~photons of scattered starlight and 1~photon of zodiacal light---assuming 1.0 local zodi and 2.0 exozodi---would be counted with each planetary photon during the exposure time $T=146,\!000$~sec that would be required to obtain a spectrum with photometric signal-to-noise ratio $\mathit{SNR}_\mathrm{pho}=10$ at the oxygen $A$~band (760 nm), as required to measure 20\% of Earth's column density of O$_{2}$ with 99\% confidence.\footnote{For details of the photometric calculation, see HandCheckPhotometry{\_}SunEarth.doc at\hfil\break http://homepage.mac.com/planetfinder/. } 

Turning to the astrometric signal, at 20~pc the amplitude of the astrometric wobble of the Sun due to the orbiting ET would be $\alpha=0.15~\mu$as or 0.00064 solar radius. $\alpha$ varies directly with the square root of the stellar luminosity and inversely with the stellar mass. An astrometric data set comprises $N$~measurements of the form ($t_i$, $\tau_i$, $x_i$, $y_i$, $\sigma_i$): the $x$--$y$ position at epochs $t_{i}$ spread over more than one planetary period, where the exposure time is $0.5\,\tau_{i}$ for each of the two quasi-orthogonal directions, and where $\sigma_i =\sigma_0 (\tau_0/\tau_i)^{1/2}$ is the positional uncertainty in either direction. The astrometric signal-to-noise ratio is $\mathit{SNR}_\mathrm{ast}=\alpha/\sigma$, where
\[
\sigma \equiv \sigma_0 \tau_0^{1/2} 
\left( {\sum\nolimits_{i=1}^N {\tau _i } } \right)^{-1/2}
=\sigma_0 \tau_0^{1/2} \tau^{-1/2}~~,
\]
where $\tau$ is the total exposure time. This noise reduction with $\tau$ and $N$ is valid only down to the noise floor, or for $\sigma>\sigma_\mathrm{floor}$, where $\sigma_\mathrm{floor}$ is determined by systematic effects. For the \textit{Space Interferometry Mission} (\textit{SIM Lite}), a space astrometer currently under study, we use the instrumental parameters are $\sigma_{0}=1.41~\mu$as, $\tau_{0}=2200$~sec, and $\sigma_\mathrm{floor}=0.035~\mu$as.\footnote{Unwin, S.~C., 
et~al. 2008, ``Taking the Measure of the Universe: Precision Astrometry with \textit{SIM PlanetQuest},'' PASP, 120, 38.}$^,$\footnote{Catanzarite, J., Shao, M., Tanner, A., Unwin, S., \& Yu, J. 2006, ``Astrometric detection of terrestrial planets in the habitable zones of nearby stars with \textit{SIM PlanetQuest},'' \textit{PASP}, 118, 1319--1339.}$^,$\footnote{Traub, W.~A.,  et~al. 2009, in \textit{Extrasolar Planets in Multi-Body Systems: Theory and Observations}, Ed. Krzysztof Gozdziewski, European Astronomical Society Publication Series, in preparation.}$^,$\footnote{Brown, R.~A. 2009, ``On the Completeness of Reflex Astrometry on Extrasolar Planets near the Sensitivity Limit,'' submitted to \textit{ApJ}; http://arxiv.org/pdf/0901.4897/.} 

Assuming periodogram (power spectrum) analysis of the data set to find weak, periodic signals, the search completeness $C(\mathit{SNR}_\mathrm{ast}$, $\mathit{fap}$) depends on $\mathit{SNR}_\mathrm{ast}$ and the accepted false-alarm probability $\mathit{fap}$; for example, $C=0.5$ for $\mathit{SNR}_\mathrm{ast}\simeq6$.
The sensitivity limit for detection probability $C=0.5$ is therefore $\alpha_\mathrm{min}=6
\times\sigma_\mathrm{floor} =0.21~\mu$as, which would be achieved in $\tau=\tau_\mathrm{max }=3.6\times 10^6$~sec. For example, an ET at 14.3~pc would be just detectable by \textit{SIM Lite} under these assumptions, because the required $\sigma = 0.21/6 = 0.035~\mu$as, which is equal to the noise floor.

\subsubsection*{The mother of all problems}

The mother of all science-operational problems for AWA is the density of stars in the solar neighborhood. If it were higher, \textit{Hubble} ($D=2.4$~m) might have obtained the first spectrum of an ET---if it were capable of $10^{-10}$ contrast, of course.

\begin{figure}[t]
\centering
\includegraphics[width=.83\textwidth]{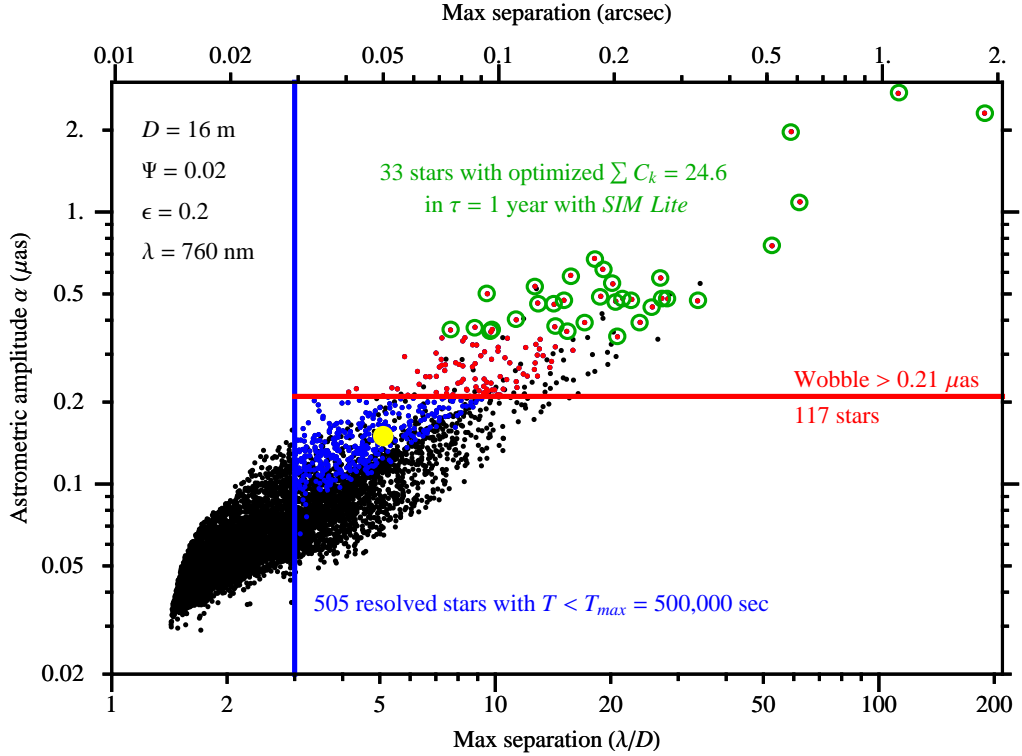}
\caption{AWA stars. Maximum apparent separations and astrometric 
amplitudes for an Earth twin (same size and temperature as Earth) around the 
7322 stars in a \textit{SIM Lite} project star list.$^{7}$ Colored stars (505): resolved at $3\,\lambda/D$ (blue line) and offering exposure time $T<500,\!000$~sec for achieving $\mathit{SNR}_{p}=10$ at $R=70$ for the $A$~band of O$_{2}$ at $\lambda =760$~nm using a 16-m telescope. Red stars (117): ET astrometric wobble $\alpha > \alpha_\mathrm{min}=0.21~\mu$as (red line), the sensitivity limit of \textit{SIM Lite} (search completeness $C=0.5$ and false-alarm probability $\mathit{fap}=0.01$). Circled red stars (33): targets in the \textit{SIM Lite} program that maximize ET discoveries (summed $C$) for 1~year of exposure time (100\% duty cycle). Yellow: Sun--Earth at 20~pc.}
\end{figure}

Figure 1 shows the 7332 stars in a \textit{SIM Lite} project star list for planet 
searching.\footnote{Thanks to Joe Catanzarite for developing and sharing this star list, which includes all the dwarfs and subgiants within 100~pc in the NStED data base (http://nsted.ipac.caltech.edu/) with stars cut by these criteria: (1)~known from the Washington Double Star Catalog to have a companion closer than 35~mas, (2)~known to have a companion between 35~mas and 1.5~arcsec with $V$~magnitude difference $<\!1$, (3)~luminosity $>\!25$, (4)~fainter than $V=9$, (5)~chromospheric activity index $R_{HK}>–4.3$, (6)~known to have age $<\!0.3$~Gyr, or (7)~with orbital period at Earth's equilibrium temperature $>\!4$ years.}
For this list, a telescope with $D=16$~m resolves 505 ETs at some time in their orbits \textit{and} achieves the required $\mathit{SNR}_{p}=10$ for the O$_{2}$ spectrum in less than $T_\mathrm{max}=500,\!000$~sec, an arbitrary, ``stretch'' exposure time. (``ET'' now means the same radius and albedo as Earth, but with semimajor axis equal to $\sqrt L $, where $L$ is the stellar luminosity, which ensures the same equilibrium temperature.) 

The exposure time for the O$_2$ spectrum for the hypothetical Earth-Sun system at 20~pc is close to the median value for the 505 stars: $T_\mathrm{median}=150,\!000$~sec. 

Some 117 of the 505 stars suitable for O$_{2}$ spectroscopy with a 16-m 
telescope have ETs detectable by \textit{SIM Lite} according to the sole criterion $\alpha > \alpha_\mathrm{min}= 0.21~\mu$as.

\subsubsection*{Finding their addresses}

For ETs, the two main searching options under consideration today are direct imaging and astrometry. Radial velocity works for larger planetary masses, but probably not ETs. For example, the 505 stars just discussed have a median mass of 0.97 and range from 0.4 to 2.4~solar masses. The radial-velocity amplitude for the Earth-Sun is  0.089~m/sec, assuming an edge-on orbit. For other stars, the ET radial-velocity amplitude varies inversely with both the square root of the mass and the fourth root of the luminosity. For the 505 stars, the median radial-velocity amplitude is 0.094 and ranges from 0.03 to 0.3~m/sec---substantially beyond current radial-velocity performance limits. Occultation photometry won't work because the alignments are highly improbable for the number of nearby stars. 

Discovery by direct detection was studied intensively for \textit{TPF-C.}\footnote{Brown, R.~A., Hunyadi, S.~L., \& Shaklan, S.~B. 2006, ``A DRM for \textit{TPF-C}: A Design Reference Mission for the Coronagraphic \textit{Terrestrial Planet Finder};''\hfil\break http://planetquest.jpl.nasa.gov/TPF/tpfcDocs/Brown.pdf/.} It tends to be ineffi\-cient---low completeness per search observation due to the highly obscured orbits, and long exposure times---but still feasible. Certainly, any telescope that could obtain the O$_{2}$ \mbox{$A$-band} spectrum of an ET at $R=70$ could detect the planet by direct imaging in a shorter time at $R =3$--10. Nevertheless, a search calls for starlight suppression over a large field of view, whereas spectroscopy requires it only at the planetary position---\textit{if the position is known.} Furthermore, direct detection has other problems that have not been solved. For example, it cannot easily determine the identity of a faint image, which could be an ET, a speckle, a background source, a cluster of exo-zodiacal light, or a companion of only indirect interest for AWA, such as a Neptune or Jupiter. The discovery image---and indeed all the data obtained in the first observing season---would provide neither the mass nor the orbit. (Solar avoidance angles are typically $>\!\!90^\mathrm{o}$, which means observing seasons $<\!\!6$ months; orbital coverage is inadequate in the first season.) First-season images could disambiguate the only a fraction of sources,  depending on a star's particular combination of parallactic ellipse and proper motion.\pagebreak\ Another debilitating consequence of not determining the orbit is that recovery of the planet in future observing seasons---predicting its position for planning and scheduling purposes---becomes a low-probability, hit-or-miss proposition.\footnote{Brown, R.~A., Shaklan, S.~B., \& Hunyadi, S.~L. 2006, ``Science Operations and the Minimum Scale of \textit{TPF},'' JPL 07-02 7/07, Proceedings of TPF-C Workshop 9/28--29/2007. (SciOpsTPF.pdf at http://homepage.mac.com/planetfinder/.)}

To further investigate astrometric discovery, we conducted a reverse auction of exposure time $\tau$ to maximize the number of planets discovered (total completeness) as a function of exposure time on \textit{SIM Lite} (assuming 100\% duty cycle). First, $\tau_{k}=\tau_\mathrm{max}$ was assigned to each of the 505 stars. Next, $\mathit{SNR}_\mathrm{ast}$ and the completenesses $C_{k}$ were computed using the ``discovery'' curve in Fig.~1 of Ref.~6. Next, those steps were repeated for $\tau_{k}$ minus 2200~sec (a quantum of exposure time), and the star was identified for which the reduction in completeness would be smallest. For that star,  the reduction in $\tau_{k}$ and associated reduction in $C_{k}$ were implemented accordingly. This process was repeated many times, always removing the smallest amount of completeness per 2200 sec of exposure time, until  the point was reached where $\Sigma\tau_k = 0$. The auction record is shown in Fig.~2. 
At $\Sigma\tau_k=1$~year, 33~stars were still being observed. For these stars, the minimum, median, and maximum values of $\mathit{SNR}_{a}$ were (5.8, 7.2, 11.1), with corresponding values of values of $C$: (0.44, 0.74, 0.99). 

\begin{figure}[t]
\centering
\includegraphics[width=.75\textwidth]{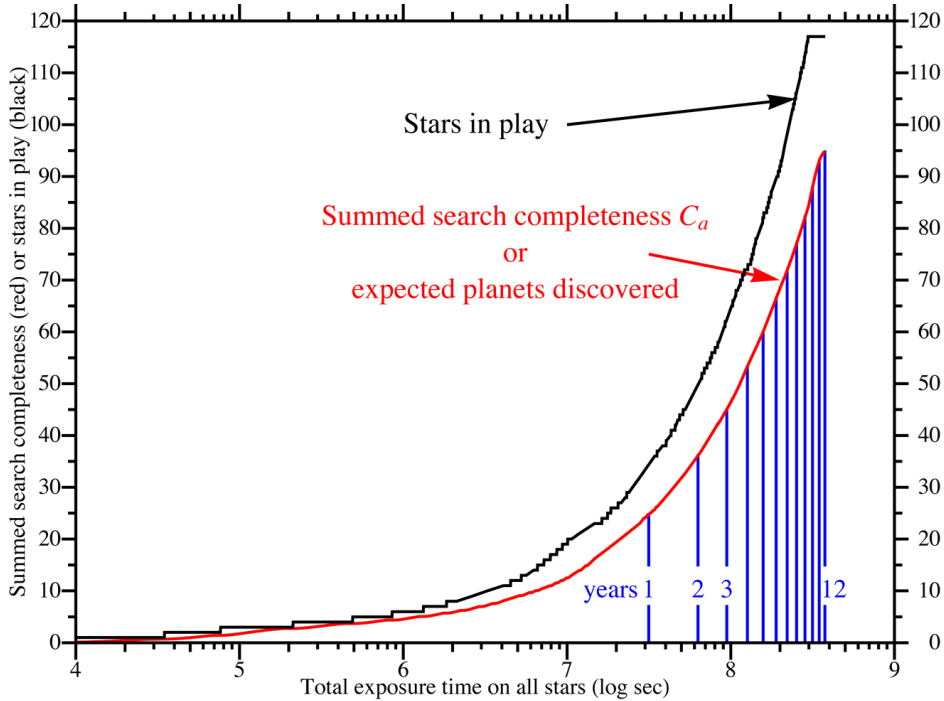}
\caption{Auction results.  Red: \textit{SIM Lite} total search completeness $\Sigma C_k$ and stars in play (still in the auction), versus total exposure time in an observing program that maximizes $\Sigma C_k$.  For 1~year total exposure time (100\% duty cycle, 33~stars are in play, and summed $\Sigma C_k=24.6$, which is the expectation value of the number of ETs discovered, assuming all stars have one. In 12~years of exposure time, all 117 stars with ETs detectable by \textit{SIM Lite} are in play, and the maximum summed completeness is achieved, $\Sigma C_k=94.7$.}
\end{figure}

The discovery of a planet---yes or no---is a Bernoulli random variable with 
probability~$C$. The expectation value of the sum of Bernoulli random variables 
is the sum of their probabilities. By summing the values of $C_{k}$, we find the expectation value of the number of ETs discovered.  For $\Sigma\tau_k=1$~year, $\Sigma C_k=24.6.$ In this case, if the occurrence rate of ETs is $\eta=0.05$, the most probable result would be one ET discovered but a 30\% chance of zero. 

For $\Sigma\tau_k=12$ years, all $\sigma=\sigma_\mathrm{min}$, all 117 stars are searched to their limits, and $\Sigma C_k= 94.7$.

Astrometry's unique forte is determining the masses and orbits of the 
ETs discovered. However, the least-squares estimators of Keplerian parameters, particularly for mass and eccentricity, are biased in the low $\mathit{SNR}_\mathrm{ast}$
regime, which is the domain of ETs for \textit{SIM Lite}.\footnote{For astrometry, see Ref.~6; for radial velocity, see: Shen, Y. \& Turner, E.~L. 2008, ``On the Eccentricity Distribution of Exoplanets from Radial Velocity Surveys,'' \textit{ApJ}, 685, 553.} Preliminary indications are that, as a result of this bias, the completeness for mass estimates as good as $\pm$25\% would be below 0.7, and below 0.3 for $\pm$±10\% accuracy, for typical ETs observed by \textit{SIM Lite}. The biased estimation of orbital elements also reduces the accuracy of positional estimates.$^{6}$

\vspace{-.1in}
\subsubsection*{Discussions}

The above calculations are \textit{science operational}. They estimate the science produced by operating the hardware. Because science is the ultimate purpose of the hardware, such calculations are essential for defining, understanding, and prioritizing exoplanetary projects.  This is particularly important for the AWA movement, both because of the dynamic, low-\textit{SNR} targets, and because complementary, concatenated  missions may be involved.

Because of unresolved science-operational issues, it is challenging to outline a coherent, low-risk AWA mission series, progressing systematically from discovery of ETs to atmospheric spectroscopy in search of biomarkers. The results of \textit{Kepler} will certainly help by clarifying what total search completeness is required to ensure an adequate sample of ETs. Even when the occurrence rate of ETs is better known, however, we still need deliberate processes for discovering, identifying, and obtaining the ephemerides of nearby ETs.

That is the top-down approach. The default, bottom-up approach would have the AWA movement not defining single-purpose missions, but proceeding opportunistically, competing for instrumental and observational opportunities on missions that are justified primarily by other science priorities. For example, the \textit{TPF-C} studies illustrate that high-performance coronagraphy is compatible with wide-field imaging---not only for optics and instruments on an off-axis, unobscured telescope, but also in terms of science operations.\footnote{R.~A.\ Brown et~al.\ 2006, ``Final Report of an Instrument Concept Study for a Wide-Field Camera for \textit{TPF-C}.'' See WFC ICS Final Report.pdf at http://homepage.mac.com/planetfinder/.} 
\newpage

Someday, \textit{Hubble}'s successor will be built to rekindle research along the long frontiers of science for which \textit{Hubble} defines the present state of knowledge. A coronagraph will surely be on board---this time with realistic hopes for the goal of ``direct imaging \ldots for planetary companions of nearby stars.''\footnote{A listed science objective in NASA's ``Announcement of Opportunity for Space Telescope,'' March 1977.} To illustratate the relative power of direct imaging, we note that all 33~stars circled in Fig.~1 are bright and close. Their ETs would be resolved with an $\mathit{IWA}=3\,\lambda/D$ for an 8-m telescope, which would achieve $\mathit{SNR}_\mathrm{pho}=10$ at $R=5$ in the $I$~band with a median exposure time of only $T=3.4\times10^3$~sec. For 27 of the 33~stars, a 4-m telescope would do the same with a median exposure time $T=5.6\times10^4$~sec. 

Similarly, in the case of space astrometry, astrophysical science other than exoplanetary was strongly endorsed by two previous decadal surveys, and it remains the \textit{SIM Lite} mission's primary justification. If \textit{SIM Lite} finds ETs, it will prove the existence theorem and produce at least rough  estimates of mass and orbit. This will be of great interest. As illustrated in Fig.~2, more observing time would reach more stars: 12~years of \textit{SIM Lite} exposure time, or 12 \textit{SIM Lite}s each exposing for 1~year, would discover all the ETs available to \textit{SIM Lite} in this star sample. 

\subsubsection*{Final remarks}

The fact we can even contemplate such challenging observations attests to NASA's progressive mastery of ever-broader wavefronts of astronomical light. This great achievement is evidenced by the success of the \textit{Hubble Space Telescope}, as well as by plans for the \textit{James Webb Space Telescope,} as well as designs for the \textit{SIM Lite} and the coronagraphic \textit{Terrestrial Planet Finder} (\textit{TPF-C}). This mastery, on which any optimistic view of the long-term future of astronomy must largely be based, will be counted among the finest achievements of the space program.

In his memoir, Riccardo Giacconi offers a comment pertinent to AWA, on ``the greatly increased program of research on extrasolar planets, stemming mainly from the personal interest of Daniel~S.\ Goldin (a manager and engineer), NASA's administrator from 1992 to 2005. This increased effort was welcomed by a minority of the astronomical community, but happened without prioritization with respect to other fields in astronomy of much greater astrophysical interest.''\footnote{Giacconi, R. 2008, \textit{Secrets of the Hoary Deep: A Personal History of Modern Astronomy}, (Baltimore: The Johns Hopkins University Press).} Hopefully, in the current decadal study, the AWA movement will be reconciled with other priorities, based on the AWA science that could actually be produced by operating the hardware, as estimated by science-operational metrics.

\acknowledgments
I thank Neill Reid, Marc Postman, Roeland van der Marel, and R\'emi Soummer for their comments, and Sharon Toolan for helping to prepare this paper.  It was originally written as a science white paper for Astro2010.

\end{document}